\documentclass[preprint,12pt]{elsarticle}

\usepackage{amssymb}

\usepackage{amsthm,amsmath,amssymb,amsopn}

\usepackage{esint}
\usepackage{graphicx,epsfig,subfig}
\usepackage{algorithm,algorithmic}
\usepackage[dvipdfmx,colorlinks,linktocpage,linkcolor=blue]{hyperref}
\usepackage{url,color}
\usepackage{threeparttable}

\def\mb{\mathbf}
\def\half{\frac{1}{2}}

\def\cal{\mathcal}
\def\bm{\boldsymbol}

\numberwithin{equation}{section}
\numberwithin{table}{section}

\journal{Computational Material Science}

\begin{document}

\begin{frontmatter}



\title{On the Cauchy-Born Approximation at Finite Temperature}


\author[whu]{Jerry Z. Yang}
\author[whu]{Chao Mao}
\address[whu]{School of Mathematics and Statistics, Wuhan University, Wuhan, Hubei, 430072, China.
}
\author[psu]{Xiantao Li}
\author[psu]{Chun Liu}
\address[psu]{Department of Mathematics, the Pennsylvania State University,
University Park, PA 16802}

\begin{abstract}
We address several issues regarding the derivation and implementation of the
Cauchy-Born approximation of the stress at finite temperature. In particular, an asymptotic
expansion is employed to derive a closed form expression for the first Piola-Kirchhoff stress. For systems
under periodic boundary conditions,  a derivation is presented, which takes into account the translational invariance
and clarifies the
removal of the zero phonon modes. Also revealed by the asymptotic approach is the role
of the smoothness of the interatomic potential. Several numerical examples are provided
to validate this approach.
\end{abstract}


\end{frontmatter}


\section{Introduction}
The recent development of molecular dynamics models have dramatically improved and enriched traditional continuum mechanics models. As a particular example,  it provides an atomistic-based constitutive
model, taking into account detailed atomic interactions. This is in contrast to many empirical models, often based
on direct observations. In addition to the desired modeling accuracy, the constitutive model derived this way automatically satisfy appropriate physical constraints, for instance, the frame indifference.

At zero temperature, the Cauchy-Born (CB) rule offers an efficient constitutive model.
The problem was first considered by Cauchy, who derived atomistic expressions
for the elastic moduli. Cauchy's work was extended by Born who also considered complex
lattices \cite{born1954dynamical}.  Given a deformation gradient, $\bm A$, the CB rule assumes that the atoms in a cell deform uniformly, from which either the strain energy density, $W_\text{CB}(\bm A)$, or the first Piola-Kirchhoff stress, $\bm P$, can be calculated.  This approximation leads to a continuum elasticity model, expressed in the
variational form
\begin{equation}
 \min \int_{\Omega}{W_\text{CB}(A)}  - \bm f(\bm x) \cdot \bm u(\bm x) d\bm x,
\end{equation}
or in the form of a PDE,
\begin{equation}
 - \nabla \cdot \bm P = \bm f(\bm x).
\end{equation}
where $W_\text{CB}(A)$ is the stored energy density.

 Thanks to the uniform deformation, the computation of $W$ and $\bm P$ can be conveniently done
in the primitive cell with periodic boundary conditions (relative to the uniform deformation)  applied. The Cauchy-Born elasticity model has been implemented in many multiscale models \cite{TaOrPh96,BeXi03,KnOr01,LiYaE09,WaLi03}, and shown great promise. The issue of the validity and accuracy of the CB rule has been addressed by several groups \cite{BlLeBrLi02,CB0,CB1,FrTh02}.
Furthermore, the CB approximation provides a fundamental link between microscopic and macroscopic descriptions of many complex physical systems, e.g., the theory of liquid crystals.  It describes the microscopic configurations under macroscopic deformations.

In principle, for systems at thermodynamic equilibrium, the CB approximation can be extended to systems at finite temperature. This, for instance, has been done in ~\cite{BlLeLePa10,Xiao2006374}. In this case, one makes the assumption that the system maintains an average deformation gradient, again denoted by $\bm A$. In addition, the fluctuation of the atomic displacement is prescribed by a Gibbs measure that corresponds to the canonical ensemble with temperature $T$. The strain energy becomes the free energy, expressed in terms of the partition function. Meanwhile, the stress can be expressed as an ensemble average with respect to the
Gibbs measure, similar to the atomic expression of the pressure \cite{BrNe95}. The continuum limit, on the other hand,  corresponds to an infinite volume limit of the Gibbs measure \cite{BlLeLePa10}.  This size-dependence, along with the probabilistic nature,  imply that the calculation has to be done over multiple cells, and multiple realizations, making the computation rather expensive, particularly when such constitutive data have to be repeatedly  accessed \cite{ChFi06,ELi05}. If one performs a standard molecular dynamics simulation to compute the stress, our experience suggests that at least a few hundred atoms have to be included in a periodic cell, and many thousand time steps are needed to equilibrate and sample the system to obtain a reasonable time average.

The primary focus of this paper is on an alternative approach to compute the stress.
In this approach, we make use of the quasi-harmonic approximation and convert the formulas to
the Fourier space -- the first Brillouin zone. This method has been used in previous works
to find closed form expressions for the free energy, e.g. \cite{Xiao2006374,DuTaMiPh05,Najafabadi1993104}. To our knowledge, however, formulas
for the stress have not been found.  More importantly, we emphasize the following issues:
\begin{itemize}
\item  Although the formula for the free energy has been previous obtained and well aware of in the community, there is a technical difficulty in deriving the formula. This difficulty can be attributed to the translational invariance of the potential energy.
As a result, the partition function is divergent. The implication to the stress formula is that the ensemble average is not well defined. Therefore, the first objective of this paper is to clarify this issue, and re-examine these formulas.

\item The quasi-harmonic approximation is typically believed to be reasonable below half of the
melting temperature \cite{AsMe76}. Outside this regime, a higher order approximation is needed. This, so far, has remained as  an open problem. The second objective of this paper is to formulate the problem as an asymptotic expansion of an integral. A systematically approach, known as the Laplace method, will be discussed.
In particular, the first order approximation is the Cauchy-Born rule at zero temperature, and the second term in the expansion agrees with that of the quasi-harmonic approximation. This points a new direction to obtain better approximations.

\item This paper brings up another important issue, regarding the role of the smoothness of the interatomic potential. We will show that the error of the quasi-harmonic approximation depends heavily on the smoothness of the empirical potential. In particular, most embedded atom potentials are parameterized using cubic spline representations, which are only $\cal{C}^2$.
Examples will show that such smoothness is insufficient.

\end{itemize}

The paper is organized as follows.  First, we discuss the derivation of the {\it exact} formula for the stress. We then
discuss the asymptotic expansion and make connections to the quasi-harmonic approximations. Emphasis is placed on the case with periodic boundary conditions, in which case the formula can be drastically simplified. In section \ref{sec: tests}, we present several numerical results for various interatomic potential models.

\section{The Finite Temperature Cauchy-Born Approximation}

This section focuses on formulas for the calculation of the average stress. More specifically, we consider an atomistic system under a uniformly applied deformation gradient, $\bm A$, and temperature $T$. We first derive a closed form expression for the exact stress, expressed as an average with respect to a canonical ensemble \cite{tadmor2011modeling,FrSm02}. Then we consider approximations with  both Dirichlet and periodic boundary conditions.

\subsection{The molecular expression of the stress}\label{sec:forward}

\begin{figure}[htbp]
\begin{center}
\includegraphics[scale=0.54]{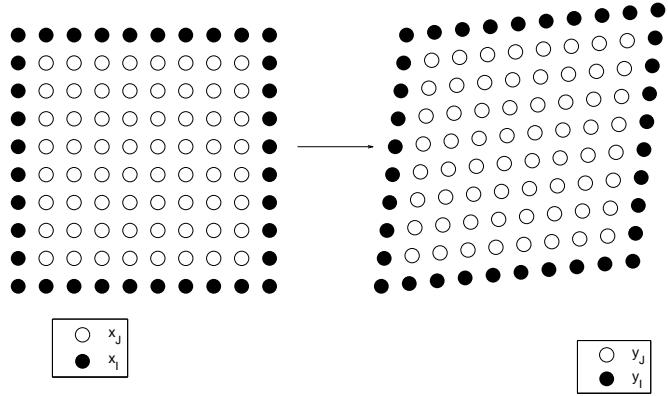}
\end{center}
\caption{An atomistic system under a uniform deformation. Left: the reference state; Right: the deformed state.
Filled circles indicate atoms along the boundary.}
\label{fig:defm}
\end{figure}

To begin, we divide the atoms into two groups: The atoms at the boundary, and the atoms inside the domain. Their reference positions are denoted by  $\bm x_J$ and $\bm x_I$, respectively; we let
$\bm x=(\bm x_I, \bm x_J)$. We also let $\bm y_J$ and $\bm y_I$ be their current position.
Figure \ref{fig:defm} shows the reference and current state of an atomistic system.

First, we consider the case where the position of the atoms at the boundary is prescribed according to a uniform deformation $\bm A$, i.e., $\bm y_I=\bm A\bm x_I$. This boundary condition is of Dirichlet type. Meanwhile, the displacement of the atoms in the interior are denoted by $\bm u_J=\bm y_J-\bm A \bm x_J$.

Let $V(\bm y_I, \bm y_J)$ be the potential energy of the system, and we sample $\bm u_J$ from the canonical ensemble,
\begin{equation}\label{eq: canon}
 \bm u_J \sim \rho_{\text{eq}},\quad \rho_{\text{eq}}=\frac{1}{Z}\exp\{- \beta V(\bm A\bm x_I, \bm A\bm x_J + \bm u_J)\}.
\end{equation}
Here $\beta= 1/(k_B T)$, $T$ is temperature, $k_B$ is the Boltzmann constant, and
$Z$ is the partition function,
\begin{equation}\label{partition}
Z= \int \exp\{- \beta V(\bm A\bm x_I, \bm A\bm x_J + \bm u_J)\} d\bm u_J,
\end{equation}
which serves as a normalizing constant.

We proceed by defining the free energy density in the canonical ensemble, given by~\cite{fecvMD,pathria1996statistical}
\begin{equation}\label{eq:freeE}
  F(\bm A, T)= -\frac{k_B T}{\Omega} \ln Z.
\end{equation}
Here $\Omega$ is the volume of the system in the reference coordinate. For example, if the interior
of the system consists of $N$ unit cells, each of which has volume $\cal{V}_0$, then
$\Omega= N \cal{V}_0$.

 Our goal is to compute the stress, given by,
 \begin{equation}\label{eq:presure}
P_{\alpha,\beta}= \frac{\partial  F(\bm A,T)}{\partial A_{\alpha,\beta}}.
\end{equation}
This is the constitutive assumption usually made in continuum mechanics models. It is also
consistent with the second law of thermodynamics.

By inserting Eq.\eqref{eq:freeE} and Eq.\eqref{partition} into Eq.\eqref{eq:presure}, we get,
\begin{equation}
\begin{aligned}
P_{\alpha,\beta} & =-\frac{k_B T}{\Omega Z} \frac{\partial  Z}{\partial A_{\alpha,\beta}}\\
& = \frac{1}{\Omega} \int \rho_{\text{eq}}(\bm u_J) \frac{\partial V}{\partial A_{\alpha,\beta}}  d\bm u_J.
\end{aligned}
\end{equation}

To simplify the expression, we define,
\begin{equation}
\phi_{\alpha,\beta}= \frac{\partial V}{\partial A_{\alpha,\beta}}
 = -\sum_{k} f_k^\alpha  x_k^{\beta},
\end{equation}
where,
\begin{equation}
f_k^\alpha= -\frac{\partial}{\partial y_k^\alpha}  V,
\end{equation}
and $y_k^\alpha$ is the $\alpha$th component of the vector $\bm y_k$.

Direct differentiation yields,
\begin{equation}\label{eq: P-avg}
P_{\alpha,\beta} = \frac{1}{\Omega} \int \rho_{\text{eq}} \phi_{\alpha,\beta} d\bm u_J = \frac{1}{\Omega}\Big\langle \phi_{\alpha,\beta} \Big\rangle.
\end{equation}
As a result, the stress is expressed as an ensemble average. This expression agrees with the virial stress in the reference coordinate~\cite{zimmerman2004calculation,tsai1979virial,citeulike:8691514}. At zero temperature, the stress is reduced to,
\begin{equation}\label{eq:0T}
P_{\alpha,\beta} = \frac{1}{\Omega} \phi_{\alpha,\beta}.
\end{equation}
Namely, no ensemble average is needed.

\subsection{An asymptotic expansion using the Laplace method}
Instead of computing the ensemble average directly, we now derive an asymptotic approximation of the stress. We first present an asymptotic approach so that the order of the error can be observed.  Notice that in the ensemble average \eqref{eq: P-avg}, both the numerator and denominator are integrals of the form,
\begin{equation}
 I(\lambda) = \int_{\mathbb{R}^d} g(\bm z) e^{-\lambda \psi(\bm z)} d\bm z.
\end{equation}
Here, $\lambda$ is assumed to be a large parameter, $\lambda\gg 1$. In the case of \eqref{eq: P-avg},
$\lambda=\beta$.

Assume that $\bm z_0$ is a local minimum of \(\psi\). Then the integral can be expanded in asymptotic series as follows (\cite{BlHa86}, Eq. (8.3.50)),
\begin{equation}
I(\lambda) =  \frac{e^{\lambda \psi(\bm z_0)} }{|\text{det}(\nabla^2 \psi(\bm z_0))|^\half}
 \left(\frac{2\pi}{\lambda}\right)^{d/2} \Big[g(\bm z_0) + \frac{1}{2\lambda} g_1 + \frac{1}{8\lambda^2} g_2 +  \cdots].
\end{equation}
The series can be obtained by a local coordinate transformation, a technique known as the Laplace method (\cite{BlHa86},
section 8.2-8.3). Intuitively, because of the exponential function, the most important contribution to the integral can only be found near a local minimum.   Explicit formulas are available for $g_j$, $j\ge 1$. In particular, we have,
\begin{equation}
 g_1 = \nabla^2 g(\bm z_0) \bm : \big[\nabla^2 \psi(\bm z_0)\big]^{-1} - \Psi_1 \nabla g(\bm z_0)
 - g(\bm z_0) \Psi_2.
\end{equation}
Here  $\Psi_1$ and $\Psi_2$ are functions that only depend on $\psi$, and the product of two matrices is defined as $L\bm:\-M= \sum_{i}\sum_j L_{ij} M_{ij}.$

Therefore, an average in the following form can be estimated,
\begin{equation}\label{eq: lap}
\begin{aligned}
&\qquad\frac{ \int_{\mathbb{R}^d} g(\bm z) e^{-\lambda \psi(\bm z)} d\bm z}
{ \int_{\mathbb{R}^d} e^{-\lambda \psi(\bm z)} d\bm z}\\
&= \frac{g_0 + \frac{1}{2\lambda} g_1 + \cal{O}(\frac{1}{\lambda^2})}
{1 + \frac{1}{2\lambda}(- \Psi_2) + \cal{O}(\frac{1}{\lambda^2})}\\
&= g_0 + \frac{1}{2\lambda}\Big( \nabla^2 g(\bm z_0) \bm : \big[\nabla^2 \psi(\bm z_0)\big]^{-1} - \Psi_1 \nabla g(\bm z_0) \Big) + \cal{O}(\frac{1}{\lambda^2}).
\end{aligned}
\end{equation}
This yields an approximation with error on the order of $\lambda^{-2}$.

\subsection{The quasi-harmonic approximation}

In this section, we consider a direct approximation: We approximate the Gibbs measure with
a Gaussian distribution.
 This can be obtained by a Taylor expansion of $V(\bm y)$ around the uniformly deformed state, yielding,
\begin{equation}
V(\bm y)\approx V(\bm A\bm x) + \half \bm u_J^T D(\bm A) \bm u_J.
\end{equation}
Here we have used $\bm A\bm x$ to indicate the uniform deformation; $D(A)= \frac{\partial^2 }{\partial \mb y_J^2} V(A\bm x)$. The linear term vanishes because of the inverse symmetry of the
Bravais lattice. If there are
$N$ particles inside, then the dimension of $D$ is $3N\times 3N$.
This approximation is commonly known as the quasi-harmonic approximation \cite{DuTaMiPh05,Najafabadi1993104}.

 This approximation of $V$ will be inserted into \eqref{eq: P-avg}, which gives an average with respect to a normal distribution, with zero average, and variance $k_B T G(\bm A)$, in which
\begin{equation}\label{eq: D2G}
G(\bm A)=D(\bm A)^{-1}.
\end{equation}

Meanwhile, we may expand the function $\phi$ in the same way,
\begin{equation}\label{eq:expansion}
\phi_{\alpha,\beta}(\bm y)\approx \phi_{\alpha,\beta}(\bm A \bm x) +
  \bm u_J^T \frac{\partial }{\partial \bm u_J} \phi_{\alpha,\beta}(\bm A \bm x)
  + \half \bm u_J^T \frac{\partial^2 }{\partial \bm u_J^2} \phi_{\alpha,\beta}(\bm A \bm x) \bm u_J.
\end{equation}

To continue, we define,
\begin{equation}\label{eq:H}
H_{\alpha,\beta}=\frac{\partial^2 }{\partial \bm u_J^2} \phi_{\alpha,\beta}(\bm A \bm x),
\end{equation}
and we notice the trivial relation,
\begin{equation}\label{eq: taylor2}
 \frac{\partial^2 }{\partial \bm u_J^2} \phi_{\alpha,\beta}(A \bm x) = \frac{\partial^3 V(\bm A\bm x)}{\partial \bm u_J^2 \partial A_{\alpha,\beta}}
  = \frac{\partial D(\bm A)}{\partial A_{\alpha,\beta}}.
\end{equation}

With a direct substitution, we finally get,
\begin{equation}\label{eq: harm0}
P_{\alpha,\beta} \approx \frac{1}{\Omega}\phi_{\alpha,\beta}(\bm A \bm x) + \frac{ k_B T}{2 \Omega} G(\bm A)\bm:H_{\alpha,\beta}(\bm A).
\end{equation}

To arrive at the formulation \eqref{eq: harm0}, we first observe that the second term
on the right hand side of \eqref{eq:expansion} disappeared after the averaging because the normal distribution has zero mean. For the last term, we can express it in a quadratic form,
\[ \half \sum_{i=1}^{3N} \sum_{j=1}^{3N} (H_{\alpha,\beta})_{i,j} u_i u_j.\]
The average of $u_i u_j $ with respect to the normal distribution is given by $k_B T G(A)_{i,j}$, which is the $(i,j)$th entry of the covariance matrix.

\medskip

Clearly, the leading term in the approximation is the average stress at zero-temperature \eqref{eq:0T}. The second term, which is linearly proportional to the temperature, serves as a correction.

More importantly, this approximation is consistent with the result from the asymptotic expansion.
In fact, in \eqref{eq: lap}, if we let $g= \phi_{\alpha,\beta}/\Omega$, then the equilibrium condition implies that $\nabla g=0$. Further, we notice that $\nabla g = H_{\alpha,\beta}/\Omega$, and
$\nabla \psi = D$. Therefore, \eqref{eq: lap} and \eqref{eq: harm0} are consistent. This also confirms that the error of the quasi-harmonic approximation is of $\mathcal{O}\big((k_BT)^2\big).$

\subsection{Further simplification}
The formula \eqref{eq: harm0} does not involve any ensemble average, which is more tractable in practice.  But the size of the matrices $G$ and $H$ can be quite large, and the matrix inversion can be rather expensive.
Here we will consider an approximation with periodic boundary conditions (PBC) applied, in which case the formula
can be greatly simplified. When the
size of the system is sufficiently large, it is expected that both types of boundary conditions will yield the same results~\cite{allen1989computer,FrSm02}.

With PBCs applied, we immediately encounter a problem.
Strictly speaking, the canonical ensemble \eqref{eq: canon} is not well defined: In order
for the function to be integrable, it has to converge to zero toward infinity, which obviously is not the case since the potential energy is invariant under translation.
To re-interpret the canonical distribution for a system with PBCs, we observe that the linear momentum of a molecular dynamics model is conserved. As a result, the center of mass remains constant. In order to appropriately define the probability density, We will explicitly impose this constraint,
\begin{equation}
 \rho_{\text{eq}}=\frac{1}{Z}\exp\{- \beta V(\bm A\bm x + \bm u)\}
 \delta(\sum_i \bm u_i).
\end{equation}
Without loss of generality, we have moved the center of mass to the origin. In addition,
we notice that due to the applied PBCs, we do not have to divide the atoms into two groups ($I$ and $J$) since the atoms outside the system are replicas of the atoms inside.

\smallskip

We now introduce the quasi-harmonic approximation. Due to the uniform deformation and the PBCs, the matrix $D$ can be greatly simplified. When arranged in a $N\times N$ block matrix form, each block is a $3\times 3$ matrix, which is a force constant matrix between two atoms. Further, under PBCs, the force constant matrix only depends on the relative position of the two atoms. Therefore, we simply denote it by $D_{i,j}= D(\bm x_i - \bm x_j)$.  In practice,
we only need to consider a few neighboring atoms for the atom at the origin.

To simplify the approximations, we take the Fourier transform,
\begin{equation}
\widehat{\bm u}(\bm\xi) = \frac{1}{\sqrt{N}}\sum_j \bm u_j e^{-i \bm\xi \cdot \bm x_j},
\end{equation}
in which the summation is over one period and $\bm \xi$ is a point in the first Brillouin zone~\cite{PhysRevB.49.16223,kittel1996introduction}. This defines an orthogonal transformation and it takes the integral to an integral with respect to the new variables $\widehat{u}(\bm\xi)$. Further, notice that
\[\widehat{\bm u}(\bm0)=\sum_j \bm u_j\] and the delta function is easily removed by setting
\[\widehat{\bm u}(\bm0)=\bm0.\]

With this transformation, the quadratic form in the normal distribution is reduced to,
\begin{equation}
\frac{1}{2} \bm u^T D \bm u= \frac{1}{2}\sum_{\bm \xi \ne \bm 0} \widehat{\bm u}(\bm \xi)^T
\widehat{D}(\bm \xi) \widehat{\bm u}(\bm \xi).
\end{equation}
Namely, we have independent normal random variables for $\bm\xi\ne\bm0$. Here,  the dependence of the force constant matrices on the deformation gradient has been suppressed. The Fourier transform, $\widehat{D}(\bm \xi)$, is known as
the dynamic matrix \cite{AsMe76}.

The rest of the derivation is similar. The matrix $G$ that appeared in the approximation is related to the lattice Green's function~\cite{Tewary1973,trinkle:014110,doi:43509,0305-4470-36-32-307}. Again we write $G$ as a $N\times N$ block matrix.
For an atom $i$ in the interior, the relation \eqref{eq: D2G} is expressed as,
\begin{equation}\label{eq: GF}
  \sum_{j} D_{i,j} G_{j,k}= \delta_{i,k}.
\end{equation}
This is precisely the definition of a lattice Green's function~\cite{hollos05_2,hollos05_1}.

For the same reason, the matrix $G$ only depends on the relative position of two atoms.
As a result, the inner product of two such matrices can be written as the sum of the inner products of each $3\times 3$ block:
\[G:H= \sum_{i=1}^N \sum_{j=1}^N G_{i,j}:H_{i,j}.\]
Due to the translational invariance, we can simplify this to,
\begin{equation}\label{eq: GH}
G:H= N\sum_{j=1}^N G_{0,j}:H_{0,j}.
\end{equation}
Therefore, we only need to consider atoms near the origin.

This is best implemented in the first Brillouin zone~\cite{AsMe76,PhysRevB.40.3616,PhysRevB.49.16223},
denoted by   $\cal{B}$. We take the Fourier transform of the Green's function,
\begin{equation}
\widehat{G}(\xi)= \frac{1}{\sqrt{N}} \sum_j G_{0,j} e^{-i\bm x_j \cdot \bm \xi}.
\end{equation}
The inverse transform is given by,
\begin{equation}\label{eq: inv}
G_j = \frac{1}{\sqrt{N}}\sum_{\bm \xi \in \cal{B}} \widehat{G}(\bm\xi) e^{i\bm x_j \cdot \bm\xi}.
\end{equation}

Substituting the inverse transform into \eqref{eq: GH}, we find that,
\begin{equation}
G:H= N \sum_{\bm \xi \in \cal{B}} \widehat{G}(\bm \xi):\widehat{H}(\bm \xi).
\end{equation}
This is the Parseval's equality for Fourier transforms.
Further, by taking the Fourier transform of \eqref{eq: GF}, we find that
\begin{equation}
\widehat{G}(\bm \xi)= \widehat{D}^{-1}(\bm \xi).
\end{equation}

Collecting terms, we finally obtain the following formula,
\begin{equation}\label{eq: harm1}
P_{\alpha,\beta} = \frac{1}{\Omega}\phi_{\alpha,\beta}(\bm A \bm x) + \frac{ k_B T}{2\Omega}  \sum_{\bm \xi \in \cal{B}, \bm\xi\ne \bm 0} \widehat{D}(\bm \xi)^{-1}:\widehat{H}_{\alpha,\beta}(\bm \xi).
\end{equation}
Compared to \eqref{eq: harm0}, we only need to compute matrix inverse and inner products for $3$-by-$3$ matrices.

The implementation of the formula \eqref{eq: harm1} is quite straightforward. It can be done as follows,
\begin{enumerate}
\item Given the deformation gradient $\bm A$, we compute the force constant matrices $D_{0,j}(\bm A)$ for atoms close to the origin;
\item Compute $\frac{d}{dA_{\alpha,\beta}} D_j(\bm A)$,  either analytically or by finite difference approximations;
\item Generate the k-points in the first Brillouin zone~\cite{PhysRevB.13.5188}, and compute the matrix inner products for each k-point.
\end{enumerate}

\section{Numerical Results}\label{sec: tests}

In canonical ensemble with PBCs, the virial stress computed from MD   \cite{ciccotti1987simulation,nose1984molecular} with volume and temporal averaging
is generally considered to be accurate. Thus we compare the stress computed from \eqref{eq: harm1} with the results from MD simulations.

When implementing the formula \eqref{eq: harm1}, one has to specify the discrete k-points in the first Brillouin zone.  We follow the widely used procedure proposed by Monkhorst and Pack \cite{PhysRevB.13.5188}, in which a uniform set of points are generated as follows:
\begin{equation}\label{eq:k-points}
{\bm k}_{n_1, n_2, n_3} = \sum_{i}^{3} \frac{2n_i-N_i-1}{2N_i} {\bm b}_i,
\end{equation}
where ${\bm b_i}$ are the reciprocal basis vectors, $n_i=1, 2,\cdots,N_i$ and $N_i$ is the number of k-points in each direction.

The quasi-harmonic type of approximation assumes that  the potential function $V(\mb y)$ is at least $\cal{C}^3$, e.g. see Eq.\eqref{eq: taylor2}. However, the EAM potential models \cite{PhysRevB.33.7983,PhysRevB.29.6443} typically used are represented using cubic spline functions, which is only $\cal{C}^2$. Therefore the accuracy will not be guaranteed. This is will be tested in our numerical experiments.


The melting temperature of Al, Cu and Fe are 933K, 1357K, and 1812K respectively. The numerical experiments here are conducted at temperature ranging from $0K$ to $500K$. We made three different choices of the deformation gradient:
{
\[
\centering
{\bm A_0} =
\left( \begin{array}{ccc}
1 & 0 & 0 \\
0 & 1 & 0 \\
0 & 0 & 1
\end{array}
 \right),\quad
 {\bm A_1} =
\left( \begin{array}{ccc}
1.01 & 0 & 0 \\
0 & 1 & 0 \\
0 & 0 & 1
\end{array}
 \right),\quad
  {\bm A_2} =
\left( \begin{array}{ccc}
0.99 & 0 & 0 \\
0 & 1 & 0 \\
0 & 0 & 1
\end{array}
 \right).
\]}
In particular, $\bm A_0$ indicates no deformation, $\bm A_1$ is a tensile strain, and $\bm A_2$ applies a compression.
These tests are conducted for three systems: Al, Cu and Fe.
\subsection{An Al system}
For the potential for Al, we use the EAM glue potential \cite{ercolessi1994interatomic}. The structure of Al is face-centered cubic (FCC), and for the MD simulation, we choose a system of size $16a_0\times16a_0\times16a_0$ with PBCs to compute the average virial stress. The lattice constant is  $a_0 =4.032\,\text{\AA}$. Further, in MD simulations, we use the No\'se-Hoover chain technique(NHC) \cite{NHC}. The time scale is \(0.052880\) pico-second, and the step size for the time integration is \(\Delta t=0.5\). For averaging over time, we make a sample every $20$ steps.

On the other hand, for the quasi-harmonic approximation,  we compute $\frac{d}{dA_{\alpha,\beta}} D_j(\bm A)$ by finite difference with step $10^{-6}$, and choose the size of k-points in the first Brillouin zone to be $32\times32\times32$.

\begin{figure}[htbp]
\centering
\begin{tabular}{cc}
\includegraphics[width=0.48\textwidth]{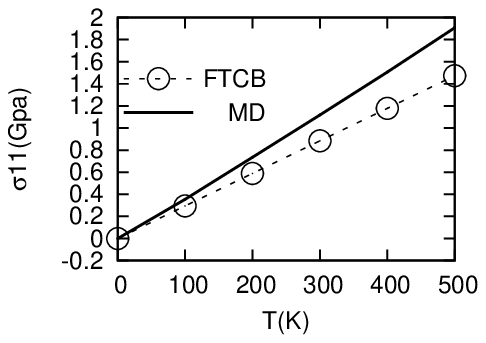}
&\includegraphics[width=0.48\textwidth]{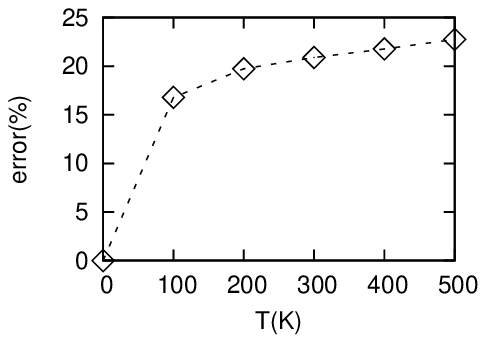}\\
\includegraphics[width=0.48\textwidth]{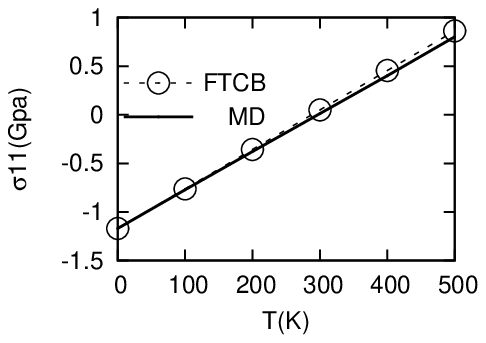}
&\includegraphics[width=0.48\textwidth]{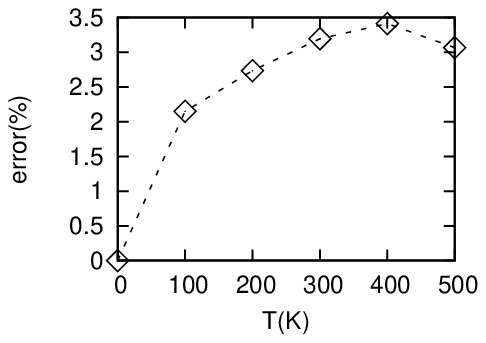}\\
\includegraphics[width=0.48\textwidth]{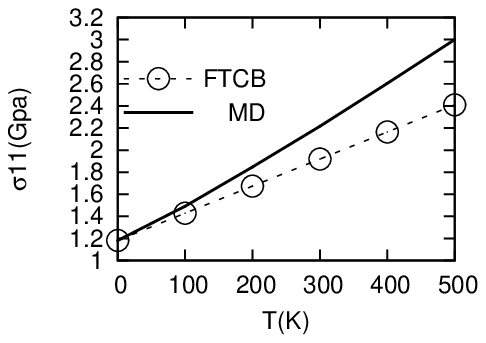}
&\includegraphics[width=0.48\textwidth]{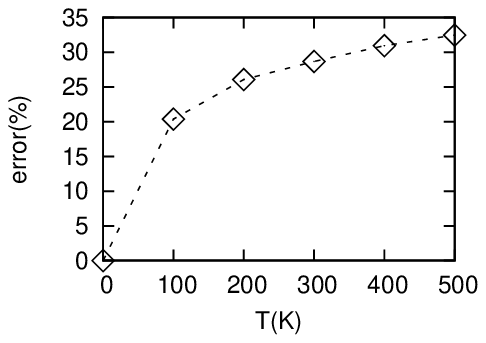}
\end{tabular}
\caption{The virial stress of the Al system computed from \eqref{eq: harm1} and MD.
From top to bottom: the deformation gradient $\bm A_0$, $\bm A_1$, and $\bm A_2$.
Right panel: The relative error. }
\label{fig:al}
\end{figure}

Figure \ref{fig:al} show the results of the approximation compared to MD. We observe that the
relative error is reasonably small for the system with a tensile strain. However the error
is considerably large for the other two cases. In particular, the error in the third case is as high as $30\%$.

\subsection{A Cu system with a smooth potential}

In this section, we present results for a Cu system.
The potential that we used is the EAM potential \cite{PhysRevB.63.224106}, given by,
\begin{equation}\label{eq: Vcu}
E = \frac{1}2\sum_{ij}V(r_{ij}) + \sum_{i}F(\bar{\rho_{i}})
\end{equation}

Here $V(r_{i,j})$ is a pair potential as a function of distance $r_{ij}$ between atoms $i$ and $j$, and $F$ is the embedding energy as a function of the host electron density $\bar{\rho_i}$, induced at site $i$ by all other atoms in the system. It is given by
\begin{equation}\label{eq: harm2}
\bar{\rho_i} = \sum_{j \neq i}\rho({r_{ij}}),
\end{equation}
where $ \rho(r) $ is the electron density function.

The pair interaction function is parametrized as follows,
\begin{equation}\label{eq:V1}
\begin{aligned}
& V(r) =- \sum_{n=1}^{3}H(r_{s}^{(n)}-r)S_{n}(r_{s}^{(n)}-r)^4\\
&+ [E_{1}M(r,r_{0}^{(1)},\alpha_{1}) + E_{2}M(r,r_{0}^{(2)},\alpha_{2}) + \delta ]\times \psi (\frac{r-r_{c}}h),
\end{aligned}
\end{equation}
where
\begin{equation}
M(r,r_{0},\alpha) = \exp[-2\alpha(r-r_{0})] - 2\exp[-\alpha(r-r_{0})],
\end{equation}
is a Morse function and $H(r)$ is the unit step function. Equation \eqref{eq:V1} includes a cutoff function $\psi(x)$ defined as $\psi(x)=0$
if $x>0$ and $\psi(x)=x^4/(1+x^4)$ if $x<0$. The last term in equation \eqref{eq:V1} is added to control the strength of pairwise repulsion
between atoms at short distances.

The electron density function takes the form:
\begin{equation}\label{eq:V2}
\rho(r)= [ a\, \exp(-\beta_{1}(r-r_{0}^{(3)})^2) + \exp(-\beta_{2}(r-r_{0}^{(4)})) ]\times \psi(\frac{r-r_{c}}h)
\end{equation}

Finally, the embedding function is represented by a polynomial:
\begin{equation}
F(\bar{\rho}) = F^{(0)} + {\frac{1}2}F^{(2)}(\bar{\rho}-1)^2 + \sum_{n=1}^{4}q_{n}(\bar{\rho}-1)^{n+2}
\end{equation}
for $\bar{\rho} < 1$ and
\begin{equation}
F(\bar{\rho}) = \frac{F^{(0)} + {\frac{1}2}F^{(2)}(\bar{\rho}-1)^2 + q_{1}(\bar{\rho}-1)^{3} + Q_{1}(\bar{\rho}-1)^{4}}{1+Q_{2}(\bar{\rho}-1)^{3}}
\end{equation}
for $\bar{\rho} > 1$. These functions have 28 parameters which can be found in Table \ref{table:Cu}.

\begin{table}
\small
\centering
\caption{Optimized values of fitting parameters of the EAM potential for Cu.}
\begin{tabular}{llll}\\
\hline
\hline
Parameter \quad  & Value \quad & Parameter \quad & Value \\
\hline
$r_c (\text{\AA})$ \quad & $5.50679$ \quad & $S_3(eV/\text{\AA}^4)$ \quad & $1.15000\times10^3$   \\
$h(\text{\AA})$ \quad & $0.50037$ \quad     & $a$ \quad & $3.80362$ \\
$E_{1}(eV)$  \quad & $2.01458\times10^2$ \quad & $r_{0}^{(3)}(\text{\AA})$ \quad & $-2.19885$ \\
$E_{2}(eV)$  \quad & $6.59228\times10^{-3}$ \quad & $r_{0}^{(4)}(\text{\AA})$ \quad & $-2.61984\times10^2$ \\
$r_{0}^{(1)}(\text{\AA})$  \quad & $0.83591$ \quad & $\beta_1 (\text{\AA}^{-2})$ \quad & $0.17394$ \\
$r_{0}^{(2)}(\text{\AA})$  \quad & $4.46867$ \quad & $\beta_2 (\text{\AA}^{-1})$ \quad & $5.35661\times10^2$ \\
$\alpha_{1}(\text{\AA}^{-1})$  \quad & $2.97758$ \quad & $F^{(0)}(eV)$ \quad & $-2.28235$ \\
$\alpha_{2}(\text{\AA}^{-1})$  \quad & $1.54927$ \quad & $F^{(2)}(eV)$ \quad & $1.35535$ \\
$\delta(\text{\AA})$  \quad & $0.86225\times10^{-2}$ \quad & $q_{1}(eV)$ \quad & $-1.27775$ \\
$r_{s}^{(1)}(\text{\AA})$  \quad & $2.24000$ \quad & $q_{2}(eV)$ \quad & $-0.86074$ \\
$r_{s}^{(2)}(\text{\AA})$  \quad & $1.80000$ \quad & $q_{3}(eV)$ \quad & $1.78804$ \\
$r_{s}^{(3)}(\text{\AA})$  \quad & $1.20000$ \quad & $q_{4}(eV)$ \quad & $2.97571$ \\
$S_{1}(eV/\text{\AA}^4)$  \quad & $4.00000$ \quad & $Q_1$ \quad & $0.40000$ \\
$S_{2}(eV/\text{\AA}^4)$  \quad & $40.00000$ \quad & $Q_2$ \quad & $0.30000$ \\
\hline
\hline
\end{tabular}
\label{table:Cu}
\end{table}

\medskip

The structure of Cu is FCC. We choose a system of size $12a_0\times12a_0\times12a_0$ with PBCs to compute virial stress. For the lattice constant, we choose $a_0 =3.803619\,\text{\AA}$.

%

\begin{figure}[htp]
\centering
\begin{tabular}{cc}
\includegraphics[width=0.48\textwidth]{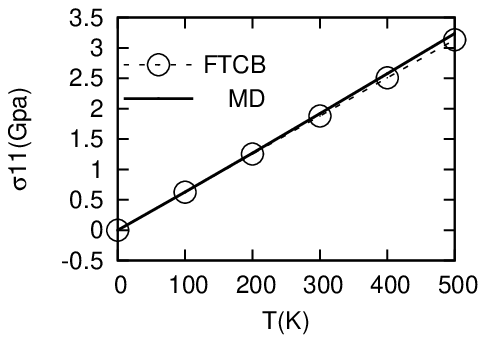}
&\includegraphics[width=0.48\textwidth]{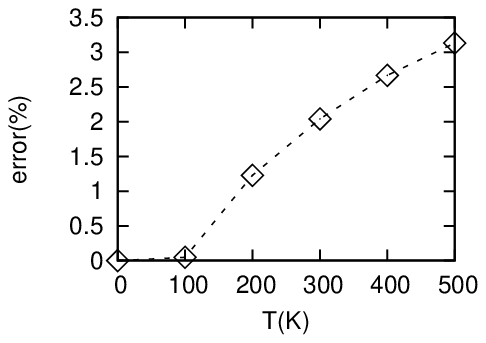}\\
\includegraphics[width=0.48\textwidth]{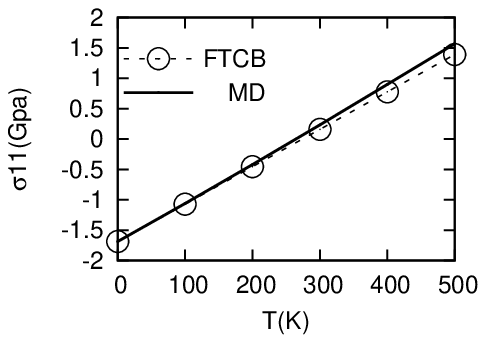}
&\includegraphics[width=0.48\textwidth]{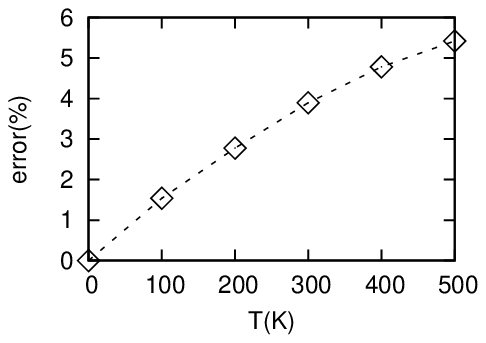}\\
\includegraphics[width=0.48\textwidth]{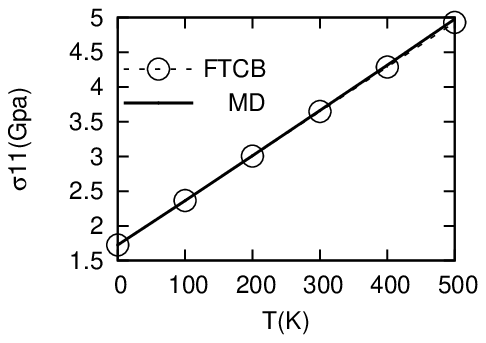}
&\includegraphics[width=0.48\textwidth]{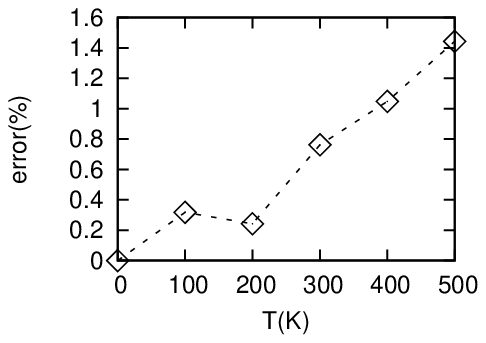}
\end{tabular}
\caption{Virial stress (\(\sigma_{11}\)) of a Cu system, computed from \eqref{eq: harm1} and MD.
From top to bottom: the deformation gradient $\bm A_0$, $\bm A_1$, and $\bm A_2$.
Right panel: The relative error.}
\label{fig:cu0}
\end{figure}

Figure \ref{fig:cu0} summarized the numerical results. The maximum relative error in all cases is less than $5.5\%$, and best results are observed in the third case, in which the relative error is less than $1.5\%$.

Both the Al and Cu systems have FCC structures, and they are both modeled by an EAM potential.
Hence, the much improved accuracy can be attributed to the smoothness of the potential.  The potential used for the Al system is represented by cubic spline functions, which is only $\cal{C}^2$, whereas  the EAM potential for Cu \eqref{eq: Vcu} is $\cal{C}^3$, and according to the asymptotic expansion, the error is expected to be smaller. In fact, since the exponential functions
in \eqref{eq:V1} and \eqref{eq:V2} decay rather quickly, we expect the jumps in the higher order derivatives to be small at the point $r_c$.

\subsection{A BCC Fe system}

We also conducted tests for a Fe system with BCC structure. Similar to the case of  Cu, we use a smoother EAM potential, first constructed in \cite{Chamati20061793}. This potential is given by
similar formulas, and the parameters are summarized in Table \ref{table:Fe}..

\begin{table}
\small
\centering
\caption{Optimized values of fitting parameters of the EAM potential for Fe.}
\begin{tabular}{llllll}\\
\hline
\hline
Parameter \quad  & Value \quad & Parameter \quad & Value \quad & Parameter \quad & Value \\
\hline
$S_{1}(eV)$ \quad & $0.5$ \quad & $r_0(\text{\AA})$ \quad & $0.50172$ \quad & $q_1 (eV)$ \quad & $-0.46026$\\
$S_{2}(eV)$ \quad & $-1.5$ \quad & $r_0^{(1)}(\text{\AA})$ \quad & $1.16319$ \quad & $q_2(eV)$ \quad & $-0.10846$\\
$S_{3}(eV)$ \quad & $0.5$ \quad & $r_0^{(2)}(\text{\AA})$ \quad & $4.70161$ \quad & $q_3(eV)$ \quad & $-0.93056$\\
$S_{4}(eV)$ \quad & $5.0$ \quad & $r_0^{(3)}(\text{\AA})$ \quad & $-1.80420\times10^2$ \quad & $q_4(eV)$ \quad & $0.577085$\\
$S_{5}(eV)$ \quad & $-10$ \quad & $r_0^{(4)}(\text{\AA})$ \quad & $-6.48409\times10^2$ \quad & $E_1(eV)$ \quad & $1.749351585\times10^4$\\
$\bar{\rho}_1$ \quad & $1.1$ \quad & $\alpha1(\text{\AA}^{-1})$ \quad & $4.50082$ \quad & $E_2(eV)$ \quad & $0.48482\times10^{-2}$\\
$\bar{\rho}_2$ \quad & $1.2$ \quad & $\alpha2(\text{\AA}^{-1})$ \quad & $2.23721$ \quad & $F_0(eV)$ \quad & $-2.1958$\\
$\bar{\rho}_3$ \quad & $1.6$ \quad & $\beta1(\text{\AA}^{-1})$ \quad & $5.7200\times10^{-3}$ \quad & $F_2(eV)$ \quad & $0.67116$\\
$\bar{\rho}_4$ \quad & $2.0$ \quad & $\beta2(\text{\AA}^{-1})$ \quad & $8.58106\times10^2$ \quad & $A$ \quad & $5.472212938$\\
$\bar{\rho}_5$ \quad & $2.5$ \quad & $\delta(\text{\AA})$ \quad & $-0.02924$ \quad & $h(eV)$ \quad & $0.59906$\\
$r_c(\text{\AA})$ \quad & $5.67337$ \\
\hline
\hline
\end{tabular}
\label{table:Fe}
\end{table}

%
%

\begin{figure}[htp]
\centering
\begin{tabular}{cc}
\includegraphics[width=0.48\textwidth]{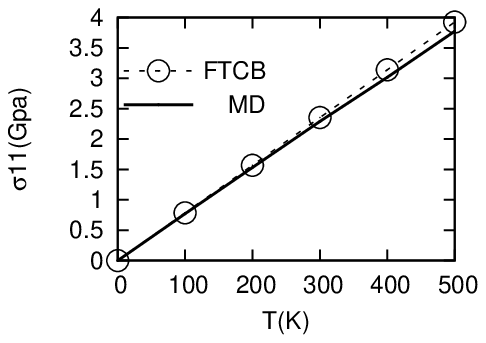}
&\includegraphics[width=0.48\textwidth]{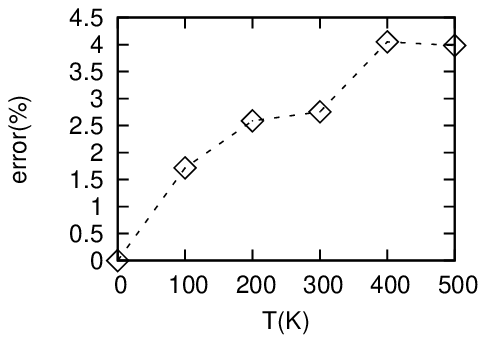}\\
\includegraphics[width=0.48\textwidth]{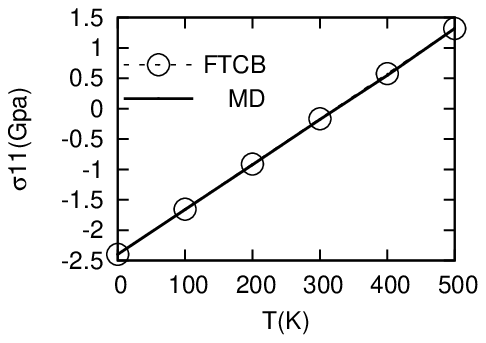}
&\includegraphics[width=0.48\textwidth]{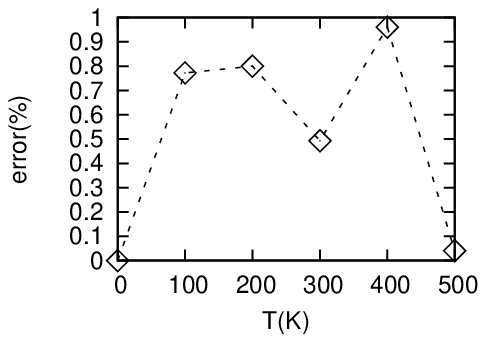}\\
\includegraphics[width=0.48\textwidth]{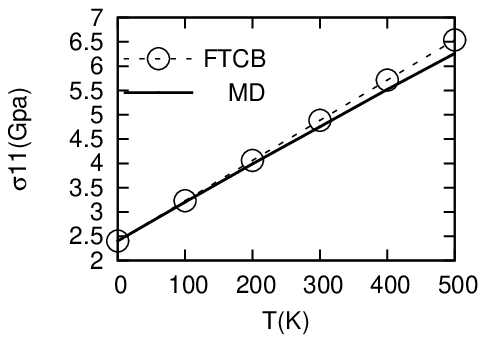}
&\includegraphics[width=0.48\textwidth]{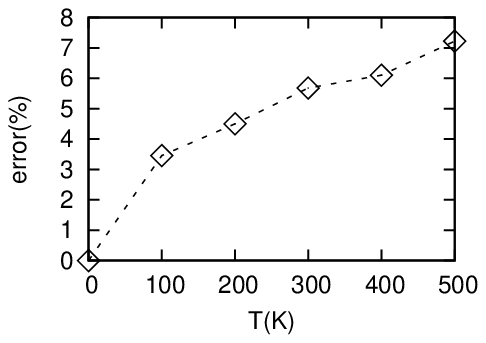}
\end{tabular}
\caption{The virial stress of the Fe system computed from \eqref{eq: harm1} and MD.
From top to bottom: the deformation gradient $\bm A_0$, $\bm A_1$, and $\bm A_2$.
Right panel: The relative error.}
\label{fig:fe0}
\end{figure}

As shown in Figure \ref{fig:cu0}, the relative error is reasonably small, and the largest error
is observed in the third case, which is under $7.5\%$, and the smallest error is observed in
the second case, where the error is less than $1.0\%$.

\begin{table}
  \centering
  \caption{CPU Time for the MD and quasi-harmonic approximation}
  \small
  \begin{threeparttable}
  \begin{tabular}{llll}\\
  \hline
  \hline
  Material\tnote{a} \quad & Structure \quad  & MD(seconds)\tnote{b}  \quad & quasi-harmonic (seconds) \\
  \hline
  Al \quad & FCC \quad  & 0.710330E+04 \quad & 0.179083E+02 \\
  Cu \quad & FCC \quad  & 0.126226E+05 \quad & 0.250343E+02 \\
  Fe \quad & BCC \quad  & 0.271006E+04 \quad & 0.970852E+01 \\
  \hline
  \hline
  \end{tabular}

     \begin{tablenotes}
     \footnotesize
       \item[a] the stress is computed at deformation $A_0$ and temperature $100K$.
       \item[b] (CPU) model name : Intel(R) Xeon(R)
       \item cpu frequency : 2.8 GB
       \item cpu memeory : 32 GB
     \end{tablenotes}
  \end{threeparttable}
  \label{table:CPUtime}
\end{table}

Table \ref{table:CPUtime} shows the CPU time for conducting one MD and quasi-harmonic calculation in the first Brillouin zone. It is worth mentioning that when the MD simulations at different temperature are performed, we have to rerun the program multiple times. In contrast,  for the quasi-harmonic approximation, due to its linear dependence on the temperature, only one implementation is sufficient.


\section{Summary and Discussion}
We have shown that the direct quasi-harmonic approximation is equivalent to the asymptotic expansion using the Laplace method \cite{BlHa86}. When periodic boundary conditions are applied, we have shown that the formula can be expressed as a sum over the k-points in the first Brillouin zone. This dramatically simplifies the implementation.

It has been mentioned in  \cite{AsMe76} that the harmonic approximation is reasonable under half of the melting
temperature. At least when the empirical potential models are smooth enough, our results confirm
that prediction to some extent. The work of Xiao and Yang \cite{Xiao2006374}  seems to offer greater accuracy, which however, may be due to the
simple Lennard-Jones potential, and the dimensions (only 1d and 2d examples were given).
 At higher temperature, this approximation breaks down; But we expect that the systematic asymptotic approach using the Laplace method has the potential to improve the accuracy. This is work in progress.




%








\end{document}